\documentstyle[11pt,newpasp,twoside,epsf]{article}
\def\edcomment#1{\iffalse\marginpar{\raggedright\sl#1\/}\else\relax\fi}
\reversemarginpar

\begin{document}
\title{Astrometry of the OH masers of 4 Mira stars}
\author{Wouter Vlemmings}
\affil{Cornell University}
\author{Huib Jan van Langevelde}
\affil{Joint Institute for VLBI in Europe}
\author{Phil Diamond}
\affil{Jodrell Bank Observatory}

\begin{abstract}
  The OH main line masers in circumstellar shells were observed with
  astrometric VLBI in order to accurately determine the position of the
  star, as well as its motion, including the parallax. Results are
  presented on 4 Mira stars monitored for 2 - 8 years with the VLBA.
  The data show that in some stars the VLBI detection is dominated by
  blue-shifted emission that is associated with the stellar image
  amplified by the maser shell in front of the star. In other cases the
  maser is not directly tied to the stellar position, but still
  persistent enough to measure proper motion and parallax.
\end{abstract}

\section{Introduction}

Accurate distances of astronomical objects are the starting point for
most astrophysical interpretations. For the optically visible Mira
variables, accurate distances have become available from the Hipparcos
astrometric mission. But a large fraction of AGB stars have so much
circumstellar material that they are too obscured to reliably measure
their motion. In addition they are Long Period Variables with large
optical amplitudes. However, some of these stars exhibit circumstellar
masers, which principle allow very accurate astrometry using VLBI.
Although H$_2$O and SiO masers yield intrinsically higher resolution,
we have concentrated on OH first. Firstly, the OH masers lie at large
distances from the star and are expected to be persistent over many
years. Furthermore, it has been proposed (Norris et al.\ 1984;
Sivagnanam et al.\ 1990) that in OH shells we can detect the amplified
stellar image, which would serve as a fixed point with respect to the
star.  Finally, astrometry at 18cm can done on the VLBA relatively
straightforwardly at the accuracy that is permitted by the OH maser
brightness ($\approx 1$ mas).

Below we describe the results of 8 year monitoring of U~Her and 2.5
years of W~Hya, S~CrB and R~Cas. Two of these stars have their VLBI
spectra dominated by blue-shifted emission, which is shown to be
consistent with the amplified stellar image. But even for the two stars
with only compact red-shifted emission it is possible to follow the
motion of the star. It is thus demonstrated that accurate distances for
enshrouded AGB stars can be obtained by VLBI. A more detailed discussion of
the results can be found in Vlemmings et al.\ (2003), early results
on U~Her were published in van Langevelde et al.\ (2000).

\section{Observations \& Results}

The 4 Mira variables were selected to have bright ($>$ 1 Jy) 1665 or
1667 (main line) maser emission and to be presumably nearby ($<$ 500
pc). Also, two bright nearby phase reference sources needed to be
available. Most of the calibrators were between $1^\circ - 2^\circ$
from the target; for S~CrB one calibrator was within $1^\circ$, for
U~Her both were at $\approx 3^\circ$. Cycle times of 5 minutes were
used, often with 2 calibrators per target in one cycle, which allows an
estimate of the accuracy.

The astrometry was obtained directly from the correlator model without
any special software, with the exception of a home grown AIPS task to
transfer calibration results from wide band data (on the reference
sources) to the 500 kHz bands on the maser. Assuming a constant
separation between pairs of extragalactic reference sources, the
precision of the relative astrometry is estimated to be $\approx 1 - 2$
mas.  Significant degradation of this accuracy was detected during the
solar maximum, as well as its dependence on ``throw'' and elevation.
In Fig.~1 \& 2 we show only the formal errors derived from the fitted
position, the residuals originate in most cases from the accuracy of
the astrometry.

\begin{figure}
\plottwo{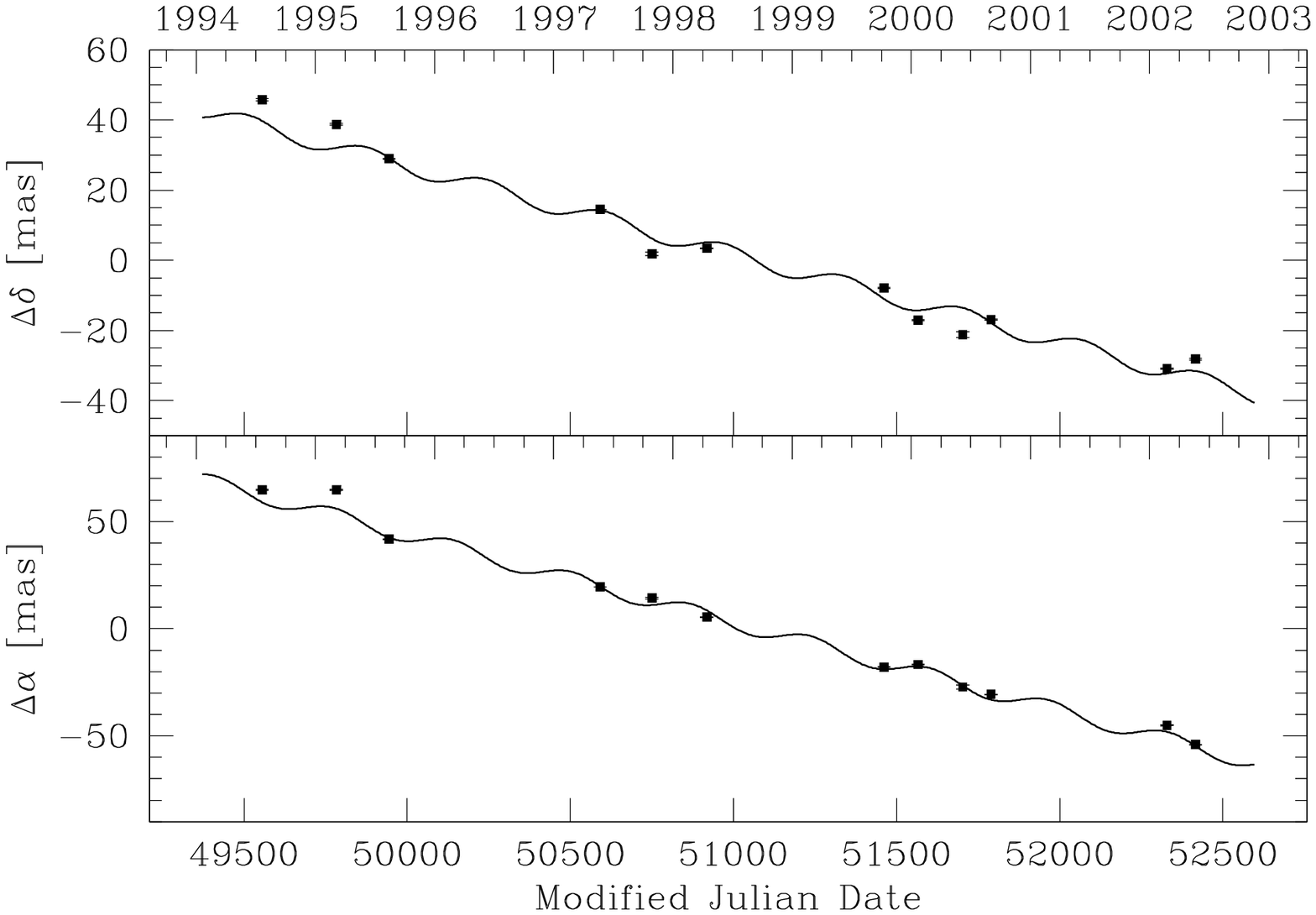}{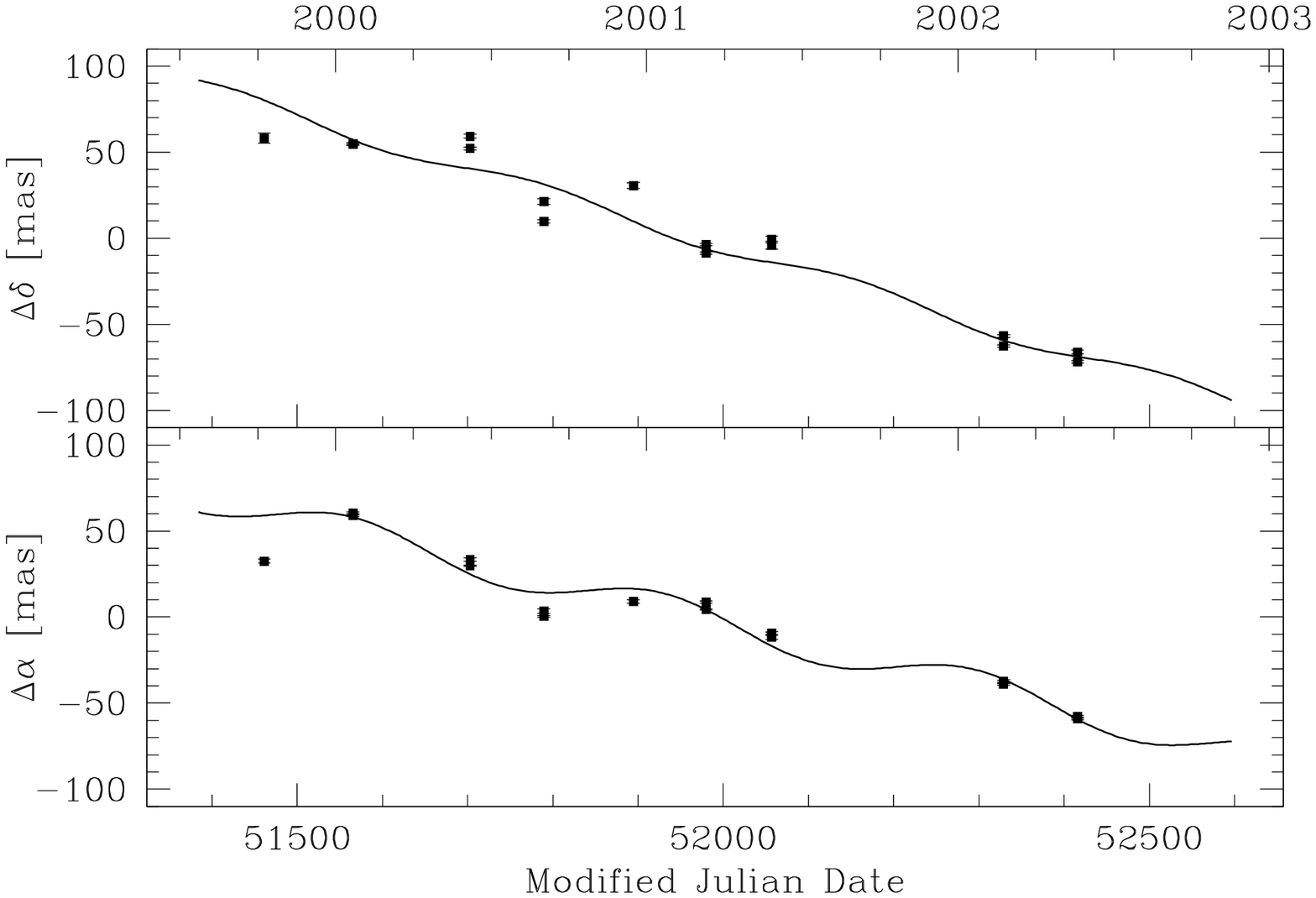}
\caption{The position of the most blue-shifted
   1667 MHz maser spot of U~Her (left) with respect to the reference
   source, spanning approximately 8 years. For W~Hya (right) both 1665
   and 1667 blue-shifted spots are used. Note that the observations
   only span 2.5 years. The drawn line represents the best fit for
   proper motion and parallax.}
\end{figure}

\begin{figure}
\plottwo{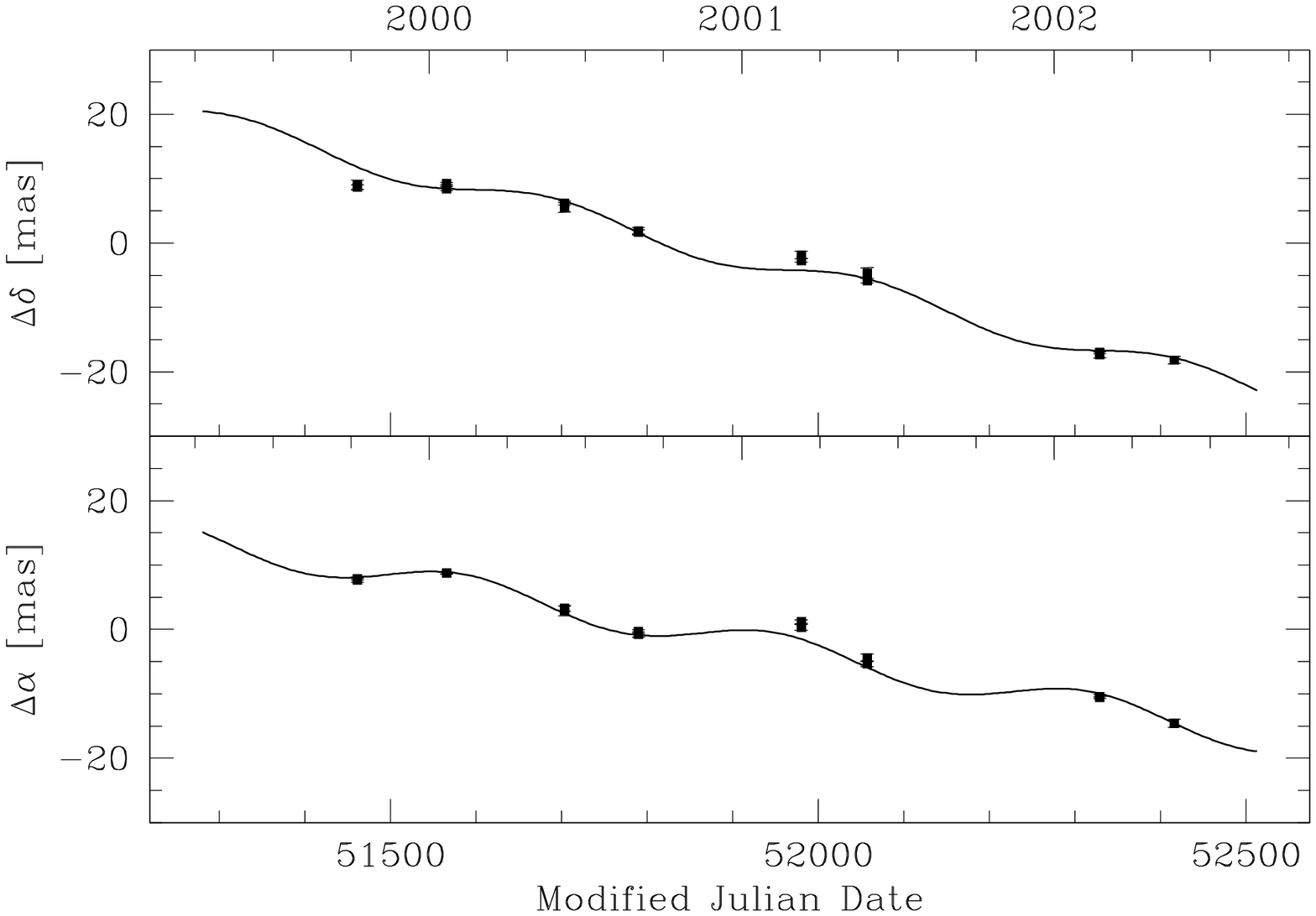}{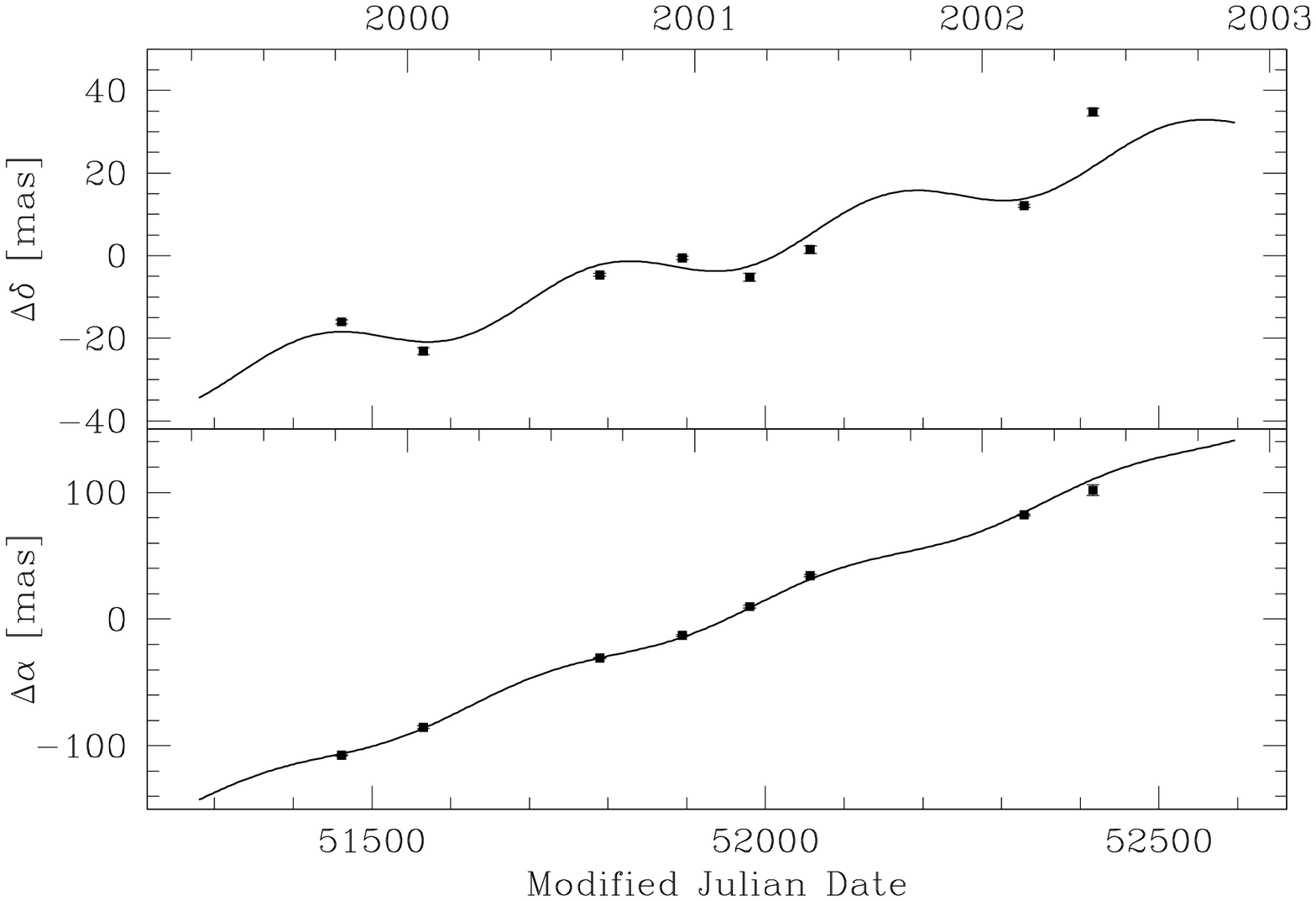}
\caption{For S~CrB (1667 \& 1665; left) and R~Cas (1665; right) only
  spots in the receding part of the shell could be found. Still,
  satisfying fits of proper motion and parallax could be made,
  especially for S~CrB which has the closest reference source.}
\end{figure}

\section{Discussion}

The relative astrometry yields both proper motions as well as
significant determinations of the parallax of each source. As these
sources are all 4 optically visible Mira stars, the proper motions can
be compared directly with results from the Hipparcos mission. In most
cases they agree within the uncertainties. 

The Hipparcos catalogue entries for the 4 target sources are flagged as
being rather unreliable, but still optical parallaxes are available,
albeit with rather large uncertainties. For S~CrB and U~Her the VLBI
results seem more precise, while the R~Cas value is significantly
different from the Hipparcos value. The VLBI value puts this star
closer to the $P-L$ relation established by (Whitelock \& Feast 2000)
For W~Hya the VLBI results seems less accurate than the Hipparcos
value; this star is very close and the residuals are larger than can be
explained from the astrometry and seem to correlate with its
variability. Maybe we are witnessing structural changes in the maser
tied to the stellar cycle.

Absolute astrometry can be done as well, because most of the reference
sources have a position determined with respect to the celestial
reference frame.  For all 4 targets this yields absolute astrometry to
about $\approx 2$ mas accuracy.  This allows a direct comparison
between the maser position and the Hipparcos optical position, where
the main limitation is set by the extrapolation of the proper motion to
a common epoch. In the cases of the blue-shifted masers these positions
agree within the errors to lie within the expected radio-sphere of the
stars.  For the red-shifted masers, the separation can be substantial.
From this we conclude that the brightest maser spots are not
exclusively produced by amplified stellar emission, but that when there
is a significant maser screen in front of the star, the line of sight
to the star is favored.

\end{document}